\documentclass[11pt,a4paper]{article}
\usepackage[T1]{fontenc}
\usepackage[utf8]{inputenc}
\usepackage{lmodern,microtype}
\usepackage[margin=27mm]{geometry}
\usepackage{amsmath,amssymb,bm,graphicx,booktabs,tabularx,authblk}
\usepackage{hyperref}
\hypersetup{hidelinks,pdftitle={From Bellissard's gap labeling to atomistic quasiperiodic bilayers},pdfauthor={Thomas D. Kuehne, Vladislav Efremkin, Emil Prodan}}
\newcommand{\Tr}{\operatorname{Tr}}
\newcommand{\tr}{\operatorname{tr}}
\newcommand{\spec}{\operatorname{spec}}
\newcommand{\ii}{\mathrm{i}}
\newcommand{\dd}{\mathrm{d}}
\newcommand{\Id}{\mathbf{1}}
\title{From Bellissard's gap labeling to atomistic quasiperiodic bilayers}
\author[1,2,3]{Thomas D. K\"uhne\thanks{Email: \href{mailto:tkuehne@cp2k.org}{tkuehne@cp2k.org}}}
\author[1,2]{Vladislav Efremkin}
\author[4]{Emil Prodan}
\affil[1]{Center for Advanced Systems Understanding, Conrad-Schiedt-Stra{\ss}e 20, 02826 G\"orlitz, Germany}
\affil[2]{Helmholtz-Zentrum Dresden-Rossendorf, Bautzner Landstra{\ss}e 400, 01328 Dresden, Germany}
\affil[3]{Institute of Artificial Intelligence, Technische Universit\"at Dresden, Helmholtzstra{\ss}e 10, 01069 Dresden, Germany}
\affil[4]{Department of Physics, Yeshiva University, New York, New York 10016, USA}

\date{}

\begin{document}
\maketitle
\begin{abstract}
Bellissard's framework uses covariant observables to describe aperiodic solids. We examine how butterfly-like spectra in twisted bilayers can be related to bulk gap labels and topological response. A bulk gap permits a spectral projector and an integrated density of states. A topological response requires an additional pairing of that projector. In an illustrative finite honeycomb bilayer, an exact second-moment identity isolates the interlayer contribution to spectral variance. Angle-resolved spectra, finite state counts, an uncoupled reference, and size and broadening checks show how coupling redistributes spectral weight without assigning a bulk gap label. We then examine the requirements for atomistic electronic-structure calculations, including nonorthogonal orbitals, projected densities of states, and basis functions that move during relative sliding. The resulting framework specifies the steps needed to connect atomistic spectra with bulk state counting and topological response.
\end{abstract}


\section{Bellissard's viewpoint beyond periodic crystals}

Bloch theory organizes the electronic structure of a periodic crystal through a family of finite matrices over the Brillouin zone. Quasiperiodic structures retain reproducible local environments without possessing a common translation cell. Their lack of a Brillouin zone does not preclude a bulk description or quantized observables. Bellissard formulated these questions using an algebra of physical observables and its $K$-theory, which organizes stable classes of projections~\cite{bellissard1986}. The noncommutative treatment of the quantum Hall effect~\cite{bellissard1994} and gap-labeling results for aperiodic patterns~\cite{bellissard2006} show how this viewpoint connects spectral and transport questions.

The key object is not an isolated finite Hamiltonian matrix. It is a covariant family of operators, together with a prescription for bulk averaging. A spectral gap then defines a projection in the observable algebra. Its class can be paired with a trace to count states and, with additional structure, with Chern cocycles to describe transport. These pairings answer related but distinct questions. In particular, a recognizable spectral pattern is neither a state count nor a measurement of a topological response.

Quasiperiodic photonic and acoustic systems make this distinction concrete. Phason-dependent pumping and boundary modes provide information beyond the density of states (DOS)~\cite{kraus2012,ni2019}. Twisted mechanical structures extend this connection to geometric incommensurability in two dimensions~\cite{rosa2021}. Electronic bilayers introduce multiple orbitals, self-consistency, structural relaxation, and nonorthogonal basis functions. We keep the same mathematical questions visible as we move from a lattice model to atomistic calculations.

This article develops that connection in three stages. We first identify the bulk objects entering gap labeling and distinguish phasons from changes of twist angle. We then use finite bilayer model calculations to separate coupling-induced spectral reconstruction from topology. Finally, we translate the operator formulation into quantities accessible to atomistic electronic-structure codes. The discussion combines established theory with an illustrative model analysis.

\section{From a family of patterns to a bulk state count}
\label{sec:bulk}

\subsection{Covariance and the trace}

Consider a lattice description with $q$ orthonormal internal orbitals per reference cell. Let $\Omega$ be a compact space of patterns with a translation action $T_{\bm n}$ for $\bm n\in\mathbb Z^2$. The unitary $U_{\bm n}$ translates lattice states according to $(U_{\bm n}\psi)_{\bm m}=\psi_{\bm m-\bm n}$. A family of bounded, self-adjoint Hamiltonians $\hat H_\omega$ on $\ell^2(\mathbb Z^2)\otimes\mathbb C^q$ is covariant if
\begin{equation}
 U_{\bm n}\hat H_\omega U_{\bm n}^{-1}=\hat H_{T_{\bm n}\omega}.
 \label{eq:covariance}
\end{equation}
Locality, continuity in the pattern, and covariance specify the observable algebra $\mathcal A$. Finite-range or sufficiently decaying hopping is a useful setting for this construction. The translation convention in Eq.~\eqref{eq:covariance} fixes the corresponding convention for the action on patterns.

For an invariant ergodic probability measure $\mu$ on $\Omega$ and a covariant observable $A=\{A_\omega\}_{\omega\in\Omega}\in\mathcal A$, the bulk trace is
\begin{equation}
 \tau(A)=\int_\Omega \tr_q\langle\bm 0|A_\omega|\bm 0\rangle\,\dd\mu(\omega)
 =\lim_{R\to\infty}\frac{\Tr(\chi_R A_\omega\chi_R)}{|\Lambda_R|}.
 \label{eq:trace}
\end{equation}
Here $\tr_q$ is the trace over the $q$ internal orbitals and $\chi_R$ selects the expanding square regions $\Lambda_R$ of reference cells. Their boundary-to-volume ratio vanishes. Under the stated ergodic assumptions, the limit holds for almost every pattern and $\tau(\Id)=q$. The normalization is states per reference cell, independent of how a DOS plot is scaled. Unique ergodicity can strengthen the independence from the chosen pattern under appropriate continuity assumptions.

Equation~\eqref{eq:trace} is a transparent lattice example, not a universal identification of a multilayer atomic structure with $\mathbb Z^2$. In a bilayer with independently varying site sets, an appropriate configuration-space or groupoid description may be required. A trace per physical area then has units of inverse area. General aperiodic frameworks address precisely the absence of a canonical global lattice indexing~\cite{bourne2018}. The choice of algebra and the normalization of its trace must precede a numerical assignment of gap labels.

\subsection{The projector and its label}

Suppose $E$ lies in a common spectral gap of the bulk family. With $\Id$ denoting the identity, functional calculus gives a spectral projection
\begin{equation}
 P_E=\Id_{(-\infty,E]}(\hat H),\qquad P_E^2=P_E=P_E^*,\qquad
 \mathcal N(E)=\tau(P_E).
 \label{eq:ids}
\end{equation}
The integrated density of states (IDS), $\mathcal N(E)$, is constant as $E$ moves within the gap. Because the discontinuity of the indicator function lies outside the spectrum, $P_E$ belongs to the observable algebra. It defines a class in $K_0(\mathcal A)$, the Grothendieck group of stable projection classes. The trace induces the homomorphism
\begin{equation}
 \tau_*:K_0(\mathcal A)\longrightarrow\mathbb R,
 \qquad [P]\longmapsto\tau(P).
 \label{eq:kzero}
\end{equation}
Bellissard's gap-labeling viewpoint constrains the possible values of the IDS through the range of this map~\cite{bellissard1986,bellissard2006}. It does not assert that every permitted label corresponds to an open gap. Nor is the trace pairing generally injective. Distinct topological classes can share a state count, so the IDS need not determine all response coefficients.

A familiar illustration is the one-dimensional quasiperiodic Harper family. For $j\in\mathbb Z$ and $\psi\in\ell^2(\mathbb Z)$, let
\begin{equation}
 (\hat H_\phi\psi)_j=t(\psi_{j+1}+\psi_{j-1})
 +V\cos(2\pi\alpha j+\phi)\psi_j,
 \label{eq:harper}
\end{equation}
Here $\alpha$ is irrational, $\phi$ is a phase modulo $2\pi$, $t$ is a nonzero real hopping amplitude, and $V$ is a real potential amplitude. There is one orbital per site. For an actual spectral gap, the trace range gives
\begin{equation}
 \mathcal N(E)=m+n\alpha\in[0,1],\qquad m,n\in\mathbb Z.
 \label{eq:harperlabel}
\end{equation}
The spectral organization is closely related to the magnetic problem underlying the Hofstadter butterfly~\cite{hofstadter1976}. For a rational approximant $\alpha=p/q$ and a gap above $r$ bands, the corresponding counting relation is $r=mq+np$. One such rational equation alone does not uniquely determine an integer pair. Continuation across the family or an independent topological calculation supplies information that an isolated finite spectrum lacks. The sign relating $n$ to a pumping convention also depends on the orientation chosen for the phason cycle.

\section{Phasons, twist, and the meaning of a response}

At fixed twist angle and layer separation, a rigid relative displacement of two ideal infinite layers changes their registry. Observed from one reference layer, the displacement modulo the other layer's lattice provides two phason coordinates. In an incommensurate setting, reference-layer translations can sample this phason torus densely. This geometric construction motivates a covariant description, but the Hamiltonian's orbital content, relaxation rule, and locality still have to respect it.

A phason cycle varies registry within a specified family. Changing the twist angle generally changes the translation action itself, while changing separation changes the strength and range of interlayer coupling. These are useful controls but are not interchangeable coordinates. In particular, an angle-dependent DOS image is not an adiabatic pumping experiment.

The distinction is explicit for a periodic approximant to a one-dimensional pump. Let $P(k,\phi)$ be a smooth occupied-band projector, gapped on the full two-torus of dimensionless $k$ and $\phi$, both of period $2\pi$. With a fixed orientation, its first Chern number is
\begin{equation}
 C=\frac{\ii}{2\pi}\int_0^{2\pi}\!\dd k\int_0^{2\pi}\!\dd\phi\,
 \tr\!\left(P[\partial_kP,\partial_\phi P]\right)\in\mathbb Z.
 \label{eq:pump}
\end{equation}
For a filled, noninteracting band subspace in the adiabatic limit, this integer determines transported particle number, with the sign set by the orientation and current convention~\cite{thouless1983}. Quasiperiodic systems implement related synthetic-dimensional responses~\cite{kraus2012,prodan2015}, and sliding moir\'e patterns provide bilayer examples~\cite{su2020,fujimoto2020}.

For suitable two-dimensional twisted models, spatial translations and two phason coordinates lead to a higher-dimensional noncommutative algebra~\cite{rosa2021}. This is a statement about the model's covariance structure, not an automatic consequence of having two atomic sheets. Extensions of gap labeling to multilayers likewise depend on the specified effective description~\cite{yoshii2023}. A response involving a spatial direction and a phason must also be distinguished from an ordinary Hall response involving two physical spatial directions. A real, time-reversal-symmetric model may have vanishing ordinary Hall conductivity while allowing a nontrivial mixed spatial-phason pump.

The spectral-gap setting is deliberately the starting point here. Quantization can also survive in a mobility gap, under suitable localization and regularity conditions~\cite{bellissard1994,bourne2018}. A mobility gap is not an interval of necessarily vanishing DOS. Inferring it requires localization information that a dark region in a broadened spectrum cannot provide.

\section{What a finite bilayer model can establish}
\label{sec:model}

\subsection{Geometric coupling}

Consider two identical honeycomb disks. Let $a$ denote the nearest-neighbor distance. The primitive vectors are $\bm a_1=(\sqrt{3}a,0)$ and $\bm a_2=(\sqrt{3}a/2,3a/2)$, with basis positions $(0,0)$ and $(0,a)$. Retain the $N$ sites $\bm r_i$ within distance $R$ of the origin. The lower disk lies at height zero. The upper disk lies at height $d$ after an in-plane rotation $\mathsf R_\theta$ through angle $\theta$. On orthonormal site states, the tight-binding (TB) Hamiltonian is
\begin{equation}
 \hat H_{\mathrm{TB}}=\sum_{i\ne j}t_0\exp(-r_{ij}/\xi)|i\rangle\langle j|,
 \qquad t_0>0,\quad \xi>0.
 \label{eq:bilayer}
\end{equation}
The sum includes every pair of distinct sites in the bilayer. Their spatial separation is $r_{ij}$. The on-site term is zero. This isotropic model does not resolve the angular dependence of carbon bonding.

The nearest-neighbor intralayer hopping $t_{\mathrm{nn}}=t_0\exp(-a/\xi)$ provides an exact energy scale. The matrix of $\hat h=\hat H_{\mathrm{TB}}/t_{\mathrm{nn}}$ is
\begin{equation}
 \bm h=\begin{pmatrix}\bm D&\bm B\\\bm B^\dagger&\bm D\end{pmatrix},
 \quad D_{ij}=(1-\delta_{ij})e^{(a-|\bm r_i-\bm r_j|)/\xi},
 \quad B_{ij}=e^{(a-\sqrt{|\bm r_i-\mathsf R_\theta\bm r_j|^2+d^2})/\xi}.
 \label{eq:block}
\end{equation}
Here $\bm D$ contains intralayer couplings and $\bm B$ interlayer couplings. The vectors $\bm r_i$ and $\bm r_j$ refer to sites of the unrotated disk. The matrices represent $\hat h$ in the specified finite site basis. The largest possible interlayer-to-intralayer hopping ratio is
\begin{equation}
 \frac{t_\perp^{\max}}{t_{\mathrm{nn}}}=e^{(a-d)/\xi}.
 \label{eq:coupling}
\end{equation}
It is attained by vertically aligned sites, not by every pair. Thus a small change of $d$ relative to the decay length $\xi$ can substantially alter hybridization. For $\xi/a=0.03$ and $d/a=0.99,0.98,0.97$, this ratio is approximately $1.40,1.95,2.72$. These dimensionless separations have no prescribed conversion to a first-principles interlayer distance.

\subsection{A trace identity and its numerical illustration}

The normalized second spectral moment of the finite matrix is
\begin{equation}
 \mu_2=\frac{1}{2N}\sum_{\nu=1}^{2N}\epsilon_\nu^2
 =\frac{\Tr\bm h^2}{2N}
 =\underbrace{\frac{\Tr\bm D^2}{N}}_{\text{intralayer}}
 +\underbrace{\frac{\|\bm B\|_F^2}{N}}_{\text{interlayer}},
 \label{eq:moment}
\end{equation}
where $\epsilon_\nu$ are dimensionless eigenvalues and $\|\cdot\|_F$ is the Frobenius norm. The first moment vanishes because the diagonal of $\bm h$ is zero, so $\mu_2$ is also its spectral variance. Equation~\eqref{eq:moment} is an exact matrix identity. It isolates the added spectral variance due to interlayer coupling without requiring a visual interpretation of a DOS map.

\begin{figure}[tbp]
 \centering
 \includegraphics[width=\textwidth]{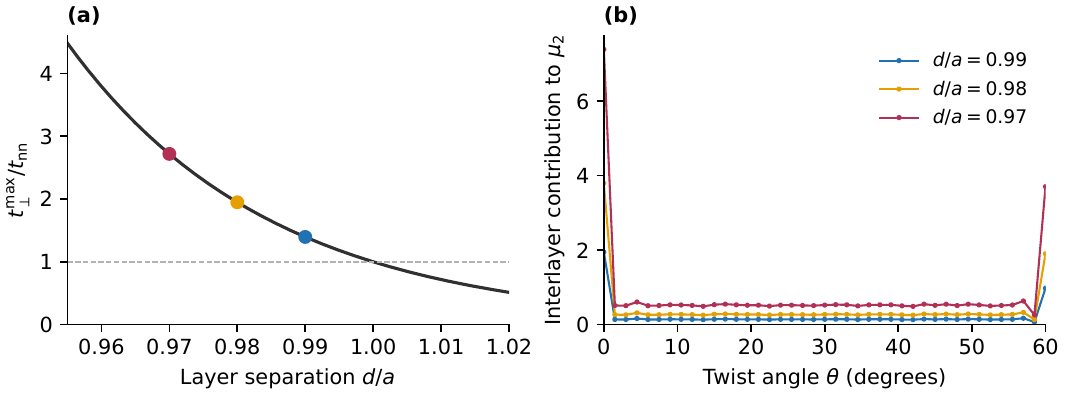}
 \caption{Geometric coupling and a finite spectral trace. (a) The maximum interlayer hopping relative to nearest-neighbor intralayer hopping, Eq.~\eqref{eq:coupling}, for $\xi/a=0.03$. Colored points mark the three calculated separations. (b) The interlayer contribution $\|\bm B\|_F^2/N$ to the second spectral moment for open honeycomb disks with $R/a=25$ and $N=1510$ sites per layer. Each curve contains 41 directly evaluated angles from $0^\circ$ to $60^\circ$, with lines connecting the samples. The same quantity is independently recovered from all 3020 eigenvalues using Eq.~\eqref{eq:moment}.}
 \label{fig:moments}
\end{figure}

Figure~\ref{fig:moments} evaluates this decomposition for $R/a=25$ and $\xi/a=0.03$. The intralayer contribution is identical for all angles and separations. The interlayer contribution depends on registry as well as separation, and increases when $d$ is reduced at fixed angle because every entry of $\bm B$ increases. Direct sums of squared hoppings and second moments of the archived eigenvalues agree to an absolute tolerance of $10^{-10}$ across all 123 spectra, validating the finite-matrix trace normalization and spectral calculation.

The relation to Bellissard's viewpoint is the distinction between a finite normalized trace and a bulk trace on a specified algebra. A second moment averages a polynomial of the Hamiltonian. Gap labeling instead concerns the trace of a spectral projection, whose existence as a stable bulk object requires a gap. A growing $\mu_2$ establishes a growing spectral variance but does not imply a new gap, a fractal spectrum, or a nonzero Chern number. Even fully converged spectral moments would not replace the projector and response calculations in Eqs.~\eqref{eq:ids} and~\eqref{eq:pump}.

The disk geometry makes the finite calculation reproducible and treats the two layers symmetrically. Its open boundaries motivate the size checks at representative registries below. Spectral information and projector information remain distinct. Equality of two DOS maps does not establish equality of their projectors, since the eigenvectors contain information that the DOS discards.

\subsection{From butterfly spectra to state counting}

\begin{figure}[!t]
 \centering
 \includegraphics[width=\textwidth]{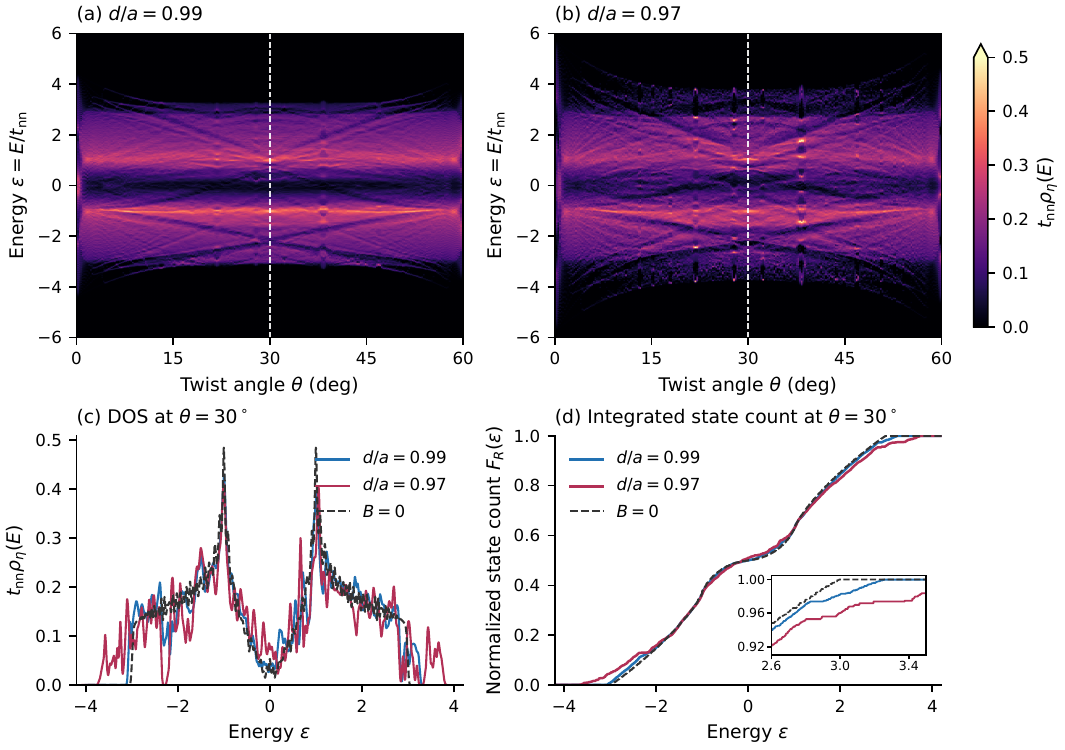}
 \caption{Butterfly-like spectral reconstruction and finite state counting in the bilayer model. (a,b) DOS maps for $d/a=0.99$ and $0.97$, with $R/a=25$, $\xi/a=0.03$, and 1510 sites in each open disk. The angle spacing is $0.25^\circ$. The DOS integrates to one and uses a common Gaussian width $\eta/t_{\mathrm{nn}}=0.02$ and color scale, clipped above $t_{\mathrm{nn}}\rho_\eta=0.5$. White dashed lines identify the $30^\circ$ cross sections. (c) DOS at that angle, without intensity clipping. (d) The exact normalized eigenvalue count, Eq.~\eqref{eq:finite-count}, with an upper-energy inset. Black dashed curves in (c,d) give the corresponding uncoupled-layer reference, $\bm B=0$, with the same geometry and per-state normalization.}
 \label{fig:butterfly}
\end{figure}

Spectral moments summarize the redistribution of eigenvalues, whereas an angle-resolved spectral map reveals its internal structure. Figure~\ref{fig:butterfly} resolves this structure at two interlayer separations using 241 angles between $0^\circ$ and $60^\circ$. The plotted DOS is the per-state Gaussian convolution
\begin{equation}
 \rho_\eta(E;\theta,d)
 =\frac{1}{2N}\sum_{\nu=1}^{2N}
 \frac{\exp[-(E-t_{\mathrm{nn}}\epsilon_\nu)^2/(2\eta^2)]}
 {\sqrt{2\pi}\,\eta},\qquad \int_{\mathbb R}\rho_\eta(E;\theta,d)\,\dd E=1,
 \label{eq:smoothed-dos}
\end{equation}
where $E$ is an energy, $\eta$ is the Gaussian standard deviation, and dependence on $R$ and $\xi$ is suppressed. The same width $\eta/t_{\mathrm{nn}}=0.02$ and color scale are used throughout. Decreasing the separation strengthens the interlayer coupling and reorganizes the curved spectral branches. These maps resolve the spectral redistribution whose second moment is given by Eq.~\eqref{eq:moment}.

The corresponding finite state count is
\begin{equation}
 F_R(\epsilon;\theta,d)=\frac{1}{2N}\sum_{\nu=1}^{2N}
 \mathbf{1}_{\{\epsilon_\nu(\theta,d)\leq\epsilon\}}
 =\frac{1}{2N}\Tr\mathbf{1}_{(-\infty,\epsilon]}(\bm h_{\theta,d}).
 \label{eq:finite-count}
\end{equation}
Here $\epsilon=E/t_{\mathrm{nn}}$, the dependence on $\xi$ is suppressed, and the normalization per site state makes $F_R$ range from zero to one. At fixed angle and separation, the DOS and $F_R$ display complementary aspects of the same eigenvalues. Spectral peaks indicate where states accumulate, while constant intervals of $F_R$ contain no finite-system eigenvalues. The $30^\circ$ cross sections in Fig.~\ref{fig:butterfly} make this connection explicit. The DOS is broadened as in Eq.~\eqref{eq:smoothed-dos}, while the state count uses the unbroadened eigenvalues.

Equation~\eqref{eq:finite-count} supplies a concrete finite-dimensional counterpart of the spectral-projector trace in Eq.~\eqref{eq:ids}. Passing to that bulk quantity entails increasing $R$, controlling the boundary contribution, and converting the per-state normalization to the cell or area normalization of the chosen trace. Bellissard's gap labels constrain the resulting bulk state count. The butterfly map, the counting function, and the convergence procedure thus play distinct, complementary roles.

\subsection{Uncoupled reference and resolution controls}

Setting $\bm B=0$ in Eq.~\eqref{eq:block} gives $\bm h_0=\bm D\oplus\bm D$ without changing the intralayer geometry or energy scale. Every monolayer eigenvalue then occurs twice, so division by $2N$ yields exactly the single-layer DOS and counting function. The reference is independent of $\theta$ and $d$. Its comparison with the coupled spectra in Fig.~\ref{fig:butterfly}(c,d) isolates interlayer-induced redistribution at a common energy origin.

We test $R/a=20,25,30$, containing $N=979,1510,2167$ sites per layer, at $\theta=0^\circ,15^\circ,30^\circ$ and both plotted separations. These angles sample aligned, intermediate, and $30^\circ$ registries. Gaussian widths $s=\eta/t_{\mathrm{nn}}=0.01,0.02,0.04$ bracket the displayed value. For each fixed $\theta$, $d$, and $\xi$, define the dimensionless DOS $\bar\rho_{R,s}(\epsilon)=t_{\mathrm{nn}}\rho_{s t_{\mathrm{nn}}}(t_{\mathrm{nn}}\epsilon;\theta,d)$. We measure size sensitivity using
\begin{equation}
 \begin{aligned}
 \mathcal W_{R,R'}&=\int_{\mathbb R}|F_R(\epsilon)-F_{R'}(\epsilon)|\,\dd\epsilon,\\
 \mathcal L_s^{R,R'}&=\int_{\mathbb R}|\bar\rho_{R,s}(\epsilon)
                         -\bar\rho_{R',s}(\epsilon)|\,\dd\epsilon.
 \end{aligned}
 \label{eq:size-distances}
\end{equation}
The first measure compares unbroadened state counts exactly. The second compares normalized DOS curves on a grid of spacing $0.0025$ over $[-7,7]$, which contains the spectra and Gaussian tails to numerical precision. The common parameters $\theta$, $d$, and $\xi$ are suppressed in Eq.~\eqref{eq:size-distances}.

\begin{table}[!ht]
 \centering
 \small
 \caption{Finite-size and broadening controls. Each entry is the maximum over the six angle--distance combinations at fixed $\xi/a=0.03$. The count distance $\mathcal W$ is in dimensionless energy units, and the DOS distance $\mathcal L_s$ lies between zero and two.}
 \label{tab:resolution}
 \begin{tabular}{@{}ccccc@{}}
 \toprule
 $R/a\to R'/a$ & $\max\mathcal W$ & $\max\mathcal L_{0.01}$ & $\max\mathcal L_{0.02}$ & $\max\mathcal L_{0.04}$ \\
 \midrule
 $20\to25$ & 0.01592 & 0.2971 & 0.1465 & 0.0872 \\
 $25\to30$ & 0.00919 & 0.2083 & 0.1114 & 0.0668 \\
 \bottomrule
 \end{tabular}
\end{table}

The size differences decrease for every sampled angle and separation, both in $\mathcal W$ and at each fixed width. Smoothing reduces the DOS difference further, as expected, so it is assessed alongside the unbroadened count rather than as a separate demonstration of convergence. At $30^\circ$, the count distance from the uncoupled reference changes from $0.04544$ to $0.04602$ for $d/a=0.99$ and from $0.13345$ to $0.13494$ for $d/a=0.97$ when $R/a$ increases from 25 to 30. The coupling-induced redistribution thus persists, while the narrowest-width DOS retains appreciable finite-size structure. A candidate bulk gap requires a persistent depleted interval under further size and boundary controls, beyond the plateaus that occur automatically between consecutive finite-system eigenvalues.

\section{From nonorthogonal matrices to physical observables}

\subsection{A change of representation is not a change of Hamiltonian}

Atom-centered basis functions $\{|\chi_\mu\rangle\}$ generally overlap. A finite-basis electronic-structure calculation solves
\begin{equation}
 \bm H\bm c_n=\varepsilon_n\bm S\bm c_n,
 \qquad H_{\mu\nu}=\langle\chi_\mu|\hat H|\chi_\nu\rangle,
 \qquad S_{\mu\nu}=\langle\chi_\mu|\chi_\nu\rangle.
 \label{eq:generalized}
\end{equation}
Choose eigenvectors normalized by $\bm c_n^\dagger\bm S\bm c_m=\delta_{nm}$. For a linearly independent basis, $\bm S$ is positive definite. Symmetric orthogonalization gives~\cite{lowdin1950}
\begin{equation}
 \widetilde{\bm H}=\bm S^{-1/2}\bm H\bm S^{-1/2},
 \qquad \widetilde{\bm c}_n=\bm S^{1/2}\bm c_n,
 \qquad \widetilde{\bm c}_n^\dagger\widetilde{\bm c}_m=\delta_{nm}.
 \label{eq:lowdin}
\end{equation}
This represents the same projected operator in an orthonormal basis. Diagonalizing $\bm H$ as though $\bm S=\bm I$ would instead solve a different problem. Near-linear dependencies must be controlled before taking inverse square roots. An overlap-eigenvalue cutoff changes the retained subspace and must be documented.

For a sharp energy cutoff in a finite-system gap, the orthonormal-coordinate projector is
\begin{equation}
 \widetilde{\bm P}_E=\sum_{\varepsilon_n\leq E}
 \widetilde{\bm c}_n\widetilde{\bm c}_n^\dagger.
 \label{eq:finiteprojector}
\end{equation}
Its ordinary matrix trace counts occupied one-particle states in that finite basis. Passing from this count to a bulk IDS still requires a controlled sequence of geometries and a fixed physical normalization. Adding Gaussian basis functions should not silently change the definition of states per cell or per area.

\subsection{Why a projected DOS is not a gap label}

Let $\bm Q_L$ be an orthogonal projector onto selected L\"owdin orbitals. The projected DOS, before broadening, is
\begin{equation}
 \rho_L(E)=\sum_n w_n^{(L)}\delta(E-\varepsilon_n),
 \qquad w_n^{(L)}=\widetilde{\bm c}_n^\dagger\bm Q_L\widetilde{\bm c}_n.
 \label{eq:projecteddos}
\end{equation}
Unlike the total state count, these weights may be fractional. If a set of mutually orthogonal orbital projectors sums to the identity, summing their DOS recovers the total DOS for the same eigenstates and normalization. A different population-analysis convention need not assign the same layer weights.

A two-level example already exposes the logical limitation. For $\widetilde{\bm H}=\operatorname{diag}(\varepsilon_a,\varepsilon_b)$ and $\bm Q_L=\operatorname{diag}(1,0)$, the projected DOS has no weight at $\varepsilon_b$, although the full Hamiltonian has an eigenstate there. More generally, $\bm Q_L\widetilde{\bm P}_E\bm Q_L$ is not a projection unless the relevant subspaces commute. Its weighted trace therefore does not inherit the interpretation of a $K_0$ gap label. A layer-resolved depletion is a useful diagnostic of orbital character, not proof of a bulk spectral gap.

Display broadening and electronic occupations require a separate distinction. Convolution with a kernel of width $\eta$ smooths spectral data and can hide small gaps. Fermi--Dirac smearing changes occupations. At temperature $T>0$, write $f_\beta(E)=[1+\exp(\beta(E-\mu))]^{-1}$ with $\beta=(k_{\mathrm B}T)^{-1}$ and chemical potential $\mu$. The matrix $f_\beta(\widetilde{\bm H})$ is generally not idempotent and is not the sharp projector appearing in Eq.~\eqref{eq:ids}. Neither operation supplies a topological classification by itself.

\subsection{Parameter-dependent bases and Berry geometry}

Orthogonalization at a fixed geometry is insufficient for a phason derivative if the basis itself moves. Let $\lambda$ parametrize a relative displacement and let $\chi(\lambda)$ map coefficient vectors to the corresponding linear combinations of physical basis functions. Write $B(\lambda)=\chi(\lambda)\bm S(\lambda)^{-1/2}$, so $B^\dagger B=\bm I$. On a local parameter patch, let the columns of $\widetilde{\bm C}(\lambda)$ form a smooth orthonormal frame for the selected states. The physical frame is $\Psi=B\widetilde{\bm C}$, and its Berry connection is
\begin{equation}
 \mathcal A_\lambda=\ii\Psi^\dagger\partial_\lambda\Psi
 =\ii\widetilde{\bm C}^\dagger\partial_\lambda\widetilde{\bm C}
 +\ii\widetilde{\bm C}^\dagger B^\dagger(\partial_\lambda B)\widetilde{\bm C}.
 \label{eq:basisconnection}
\end{equation}
The second term accounts for basis motion. Omitting it is justified only when the chosen representation makes it vanish or when an equivalent covariant overlap construction already includes it. Discrete overlaps between physical states at neighboring geometries offer a practical route, provided the cross-geometry basis overlaps are treated consistently. A basis-invariant spectrum at every geometry does not, by itself, guarantee a basis-consistent Berry curvature across geometries.

Electronic-structure packages such as CP2K provide atomistic Hamiltonians and state-resolved quantities~\cite{cp2k2020,cp2k2026}. The basis-set and pseudopotential protocol of Ref.~\cite{uzh2026} illustrates why their errors should be assessed separately. These controls remain necessary when the target quantity is topological. They do not establish that the self-consistent effective Hamiltonian defines the required covariant, smoothly parameterized family. Consistent spin treatment, structural relaxation, and branch selection of the self-consistent solution are additional requirements. A Kohn--Sham spectral gap must also be distinguished from a many-body excitation gap.

\subsection{A practical CP2K development route}

The authors' current CP2K development workflow makes the moving-basis construction concrete. It exports electronic states and evaluates physical overlaps between neighboring geometries using Gaussian basis integrals, including the nonorthogonal metric, displaced centers, and reciprocal-boundary phases. Native Wilson-loop analysis and the separate \texttt{topology\_phasons} postprocessor then provide a route to first-Chern pairings on closed two-parameter meshes and second-Chern-character pairings on four-parameter meshes, following lattice-gauge formulations~\cite{fukui2005,mochol2019}. For sliding bilayers, candidate parameter spaces are $(k_x,\phi_x)$ and $(k_x,k_y,\phi_x,\phi_y)$, respectively. This describes a development implementation, not a new material result in this article.

Such calculations require a physically closed family with consistent cell, atom count, basis, spin convention, and isolated subspace. One must check the gap, overlap conditioning, boundary sewing, and mesh refinement. In particular, a finite-mesh second-Chern estimate cannot simply be rounded to an integer. The workflow does not add a magnetic field or turn an angle-dependent DOS map into a topological classification. It enables a focused atomistic phason calculation to follow the spectral diagnostics presented here. A converged graphene invariant remains outside the scope of this perspective.

\section{A hierarchy of evidence for atomistic gap labeling}

The mathematical stability statement provides a useful standard for numerical practice. Let $E$ lie a distance $\delta>0$ from the spectrum of a bounded self-adjoint $\hat H$ in a fixed observable algebra. If a self-adjoint perturbation $\Delta\hat H$ in that algebra satisfies $\|\Delta\hat H\|<\delta$, then
\begin{equation}
 \hat H_s=\hat H+s\Delta\hat H,\quad 0\leq s\leq1,
 \qquad \operatorname{dist}(E,\spec\hat H_s)
 \geq\delta-\|\Delta\hat H\|>0.
 \label{eq:stability}
\end{equation}
The spectral projectors form a continuous path and retain their $K_0$ class. The spectral bound follows from the resolvent estimate, or equivalently from the Neumann-series criterion for invertibility. This statement explains robustness but does not turn an observed finite gap into a bulk one. Comparisons between changing basis spaces need a common embedding before an operator-norm estimate has this meaning.

Table~\ref{tab:evidence} separates the principal observables. Assigning an integer before establishing the relevant bulk projector reverses the logic of gap labeling. Conversely, demanding a quantized response from a diagnostic model moment asks that observable to answer a question it was not designed to address.

\begin{table}[!ht]
 \centering
 \small
 \caption{Different numerical objects support different physical conclusions.}
 \label{tab:evidence}
 \begin{tabularx}{\textwidth}{@{}>{\raggedright\arraybackslash}p{0.23\textwidth}>{\raggedright\arraybackslash}X>{\raggedright\arraybackslash}X@{}}
 \toprule
 Object & Direct information & Additional requirement for topology \\
 \midrule
 Finite spectral moment & Averaged spectral scale and coupling contributions & A bulk limit and a gapped spectral projector \\
 \addlinespace
 Layer-projected DOS & Spectral weight in a chosen subspace & Total spectral information and a projection-independent gap test \\
 \addlinespace
 Bulk IDS plateau & State count in an established spectral gap & A justified trace range and, where needed, further topological pairings \\
 \addlinespace
 Phason-dependent projector & Geometry of occupied states across a family & A converged invariant and the assumptions connecting it to response \\
 \bottomrule
 \end{tabularx}
\end{table}

A practical analysis begins by specifying the family of infinite structures, its covariance, the retained orbitals, and the trace normalization. Finite patches or periodic approximants can then be chosen to approach that family. For a two-dimensional open patch, boundary state counts typically grow with perimeter while bulk counts grow with area. Boundary weight can vanish in a normalized DOS and still encode physically important spectral flow. Interior projections help diagnose bulk gaps, while boundary-resolved calculations address edge behavior.

The next step is to converge the full spectral information with respect to size, approximant order, basis quality, and the relevant numerical cutoffs. Candidate gaps must be distinguished from missing calculations, insufficient energy resolution, and finite-size level spacings. A consistent energy reference is needed to follow the same gap across a parameter family. Aligning each spectrum independently to its own Fermi level can obscure that tracking.

Only after this stage should an IDS plateau be compared with the trace range of the chosen algebra. Its persistence should be checked along the appropriate gapped family. If a pumping or Hall interpretation is sought, one should compute the corresponding pairing of the projector with spatial and/or phason derivatives, using consistent parameter-dependent bases. Boundary spectral flow or an explicit adiabatic transport calculation then offers an independent test of the proposed response. These checks complement one another and cannot be replaced by a visual resemblance to the Hofstadter butterfly.

\section{Perspective}

Bellissard's work turns the interpretation of quasiperiodic spectra into a constructive program. One establishes the bulk spectral object, counts its states with the appropriate trace, and computes the pairing that describes the desired response. The bilayer calculations illustrate how this program begins. An exact moment identity identifies the interlayer contribution to spectral variance. The uncoupled reference isolates its energy-resolved redistribution, while size and broadening controls assess the resolution of that observation. Together, these diagnostics connect geometry to spectral reconstruction before an integer label is assigned.

The further passage to atomistic quasiperiodic bilayers is demanding but well posed. Nonorthogonal basis sets require metric-aware projectors, moving atoms require basis-consistent parameter derivatives, and self-consistency requires a reproducible covariant branch of solutions. Relaxation can change the pattern space itself, so its effect is not always captured by adding a small perturbation to a fixed Hamiltonian. These are opportunities for a closer interaction between electronic-structure methods and noncommutative geometry rather than reasons to separate the subjects.

The central distinction is between spectral reconstruction, gap labeling, and response. The finite-model calculations establish the first diagnostic steps. The trace and projector formulation identifies what additional information the latter two require. The CP2K development workflow supplies a route to the moving-basis overlaps needed for an atomistic extension. A focused next application would follow one candidate gap through a physically closed sliding cycle and test the associated first-Chern pairing before attempting a four-parameter response. Such material-specific calculations remain to be performed. Bellissard's framework thereby guides the passage from geometry to state counting and response.

\paragraph{Acknowledgment.}
This perspective is dedicated to Jean Bellissard on the occasion of his eightieth birthday and to the continuing influence of his work on the mathematics and physics of aperiodic solids.

\paragraph{Data and code availability.}
The source of this article, the bilayer eigenvalues, the spectral-moment and butterfly data, and the uncoupled-reference and resolution controls are available in the \href{https://github.com/DCM-Uni-Paderborn/Hofstadter-Butterfly}{project repository}. Its analysis scripts reproduce Fig.~\ref{fig:moments}, Table~\ref{tab:resolution}, the uncoupled reference, and Fig.~\ref{fig:butterfly}. The CP2K implementation discussed above refers to development revision \texttt{b717751}, documented separately in the manuscript archive. No atomistic topological invariant is reported here.

\paragraph{Use of artificial intelligence tools.}
During the preparation of this work, the authors used OpenAI ChatGPT and Codex to support language editing, mathematical exposition, the development of data-analysis and plotting scripts, consistency checks, and LaTeX preparation. The authors reviewed and edited the resulting content and take full responsibility for the article.

\bibliographystyle{plain}
\bibliography{references,bellissard_references}
\end{document}